\documentclass[aps,prd,notitlepage,twocolumn,
nofootinbib,showpacs]{revtex4-1}
\usepackage{epsfig,subfigure,graphics,color,graphicx,amsmath}

\usepackage{xcolor}
\usepackage{color}
\usepackage{epsf}
\usepackage{soul}
\usepackage{etex}
\usepackage{morefloats}

\colorlet{darkgreen}{green!50!black}
\colorlet{brightyellow}{yellow!75!red}
\colorlet{orange}{red!50!yellow}
\colorlet{darkblue}{blue!60!black}
\colorlet{darkred}{red!80!black}

\begin{document}
\title{New form of kernel in equation for  Nakanishi function}
\author{V.A. Karmanov}
\email{karmanovva@lebedev.ru}
\affiliation{Lebedev Physical Institute, Leninsky Prospekt 53, 119991 Moscow, Russia}

\date{\today}

\bibliographystyle{unsrt}
\begin{abstract}
The Bethe-Salpeter amplitude $\Phi(k,p)$ is expressed, by means of the Nakanishi integral representation, 
via a smooth function $g(\gamma,z)$. 
This function satisfies a canonical equation  $g=Ng$. 
However, calculations of the kernel $N$ in this equation, presented  previously, were restricted to one-boson exchange and, depending on method, dealt with complex multivalued functions. 
Although these difficulties are surmountable, but in practice, they complicate finding the unambiguous result.

In the present work,  an unambiguous expression for the kernel $N$ in terms of real functions is derived. For  the one-boson scalar exchange, the explicit formula for $N$  is found. With this equation and kernel, the binding energies, calculated previously, are reproduced. Their finding, as well as calculation of the Bethe-Salpeter amplitude in the Minkowski space, become not more difficult than in the Euclidean one. The method can be generalized to any kernel given by irreducible Feynman graph. This generalization is illustrated by example of the cross-ladder kernel. 
\end{abstract}
\maketitle

\section{Introduction}\label{intro}
One of the efficient methods of solving the Bethe-Salpeter (BS) equation \cite{bs}
\begin{eqnarray}\label{bse}
\Phi(k,p)&=&\frac{i^2}{\left[(\frac{p}{2}+k)^2-m^2+i\epsilon\right]
\left[(\frac{p}{2}-k)^2-m^2+i\epsilon\right]}
\nonumber\\
&\times&\int \frac{d^4k'}{(2\pi)^4}iK(k,k',p)\Phi(k',p),
\end{eqnarray}
directly in Minkowsky space,  is based
on using the Nakanishi representation \cite{nak63}  for the BS amplitude:
\begin{eqnarray}\label{bsint}
\Phi(k,p)&=&-i\int_{-1}^1dz'\int_0^{\infty}d\gamma'
\\
&\times&
\frac{g(\gamma',z')}{\left[\gamma'+m^2
-\frac{1}{4}M^2-k^2-p\cdot k\; z'-i\epsilon\right]^3}. 
\nonumber
\end{eqnarray}
(here $m$ is the constituent mass,  $M$ is the bound state mass)
and finding the function $g(\gamma,z)$ from corresponding equation. The latter equation is derived from the BS one after substituting (\ref{bsint}) into (\ref{bse}). An advantage of this method is in the fact that the function $g(\gamma,z)$  satisfies an equation which is relatively easy solvable and its solution
is smooth. Having found it numerically, one can restore by Eq. (\ref{bsint}) the BS amplitude in Minkowski space. 
Then,
 after substituting $\Phi(k,p)$, e.g., in the expression for the electromagnetic form factor one can integrate analytically over the relative momentum $k$ and express form factor in terms of the solution $g(\gamma,z)$ \cite{ckm_2009,cks_epjc}. In this way, the Minkowski BS amplitude $\Phi$ appears  as an 
auxiliary quantity which in the final result is replaced by $g$.

For the massless ladder kernel (Wick-Cutkosky model \cite{wick,cutk}) the $\gamma$-dependence of the function $g(\gamma,z)$ is reduced to the delta-function $\delta(\gamma)$ and to finite sum of its derivatives so the integral (\ref{bsint}) turns into sum of the one-dimensional ones with the functions  
$g_n(z)$, satisfying a system of equations. In non-relativistic limit, the Coulombien spectrum and wave functions are reproduced.

For massive ladder exchange kernel:
 \begin{equation} \label{K}
K(k,k')= -\frac{16\pi m^2 \alpha}{(k-k')^2-\mu^2+i \epsilon}
\end{equation}
($\alpha$ is the coupling constant which appears in the corresponding Yukawa potential $V(r)=-\frac{\alpha}{r}e^{-\mu r}$)
this method was proposed and an equation for $g(\gamma,z)$ was firstly obtained in \cite{KusPRD95}.  Subsequent researches
\cite{bs1,bs2,FrePRD14,nak2}, including the present one, were aimed to deriving, step by step, most simple forms of this equation and its kernel, convenient for practical use, further simplifying the numerical solutions.
\

In Ref. \cite{bs1}, by substituting $\Phi(k,p)$ in the form (\ref{bsint}) into Eq. (\ref{bse}) and making  the light-front projection, the following equation for the function $g(\gamma,z)$ was derived: 
\begin{eqnarray} \label{bsnew}
&&\int_0^{\infty}\frac{g(\gamma',z)d\gamma'}{\Bigl[\gamma'+\gamma +z^2 m^2+(1-z^2)\kappa^2\Bigr]^2}
=
\nonumber\\
&&\int_0^{\infty}d\gamma'\int_{-1}^{1}dz'\;V(\gamma,z,\gamma',z') g(\gamma',z'),
\end{eqnarray}
where $\kappa^2=m^2-\frac{1}{4}M^2$. 
The  kernel $V(\gamma,z,\gamma',z')$ was expressed via the  kernel $K(k,k')$, eq. (\ref{K}),  of the BS equation (\ref{bse}) in \cite{bs1}. 

The equation in the form (\ref{bsnew}), although solvable numerically,  was not yet convenient enough, since it contained the integrals in both parts. 
This resulted in instability of the numerical procedure which though was overcome in Ref. \cite{bs1} but required extra methods and studies.

 Further efforts in this field were then directed to derivation of equation in the "normal" form:
\begin{equation} \label{L6}
g(\gamma,z)=\int_0^{\infty}d\gamma'\int_{-1}^{1}dz'\;N(\gamma,z,\gamma',z') g(\gamma',z'),
\end{equation}
which does not contain the integral in l.h.-side.  It was firstly derived in \cite{FrePRD14}, using uniqueness of the Nakanishi representation. This derivation was real "matter of art", however restricted to the ladder kernel only. The binding energies found in \cite{FrePRD14} by numerical solution of Eq. (\ref{L6}) were in very good agreement with the solutions of Eq. (\ref{bsnew}) found in \cite{bs1} and with the Euclidean space solutions. Besides, the stability of results  was much better.

Then in Ref. \cite{nak2} it was noticed that the integral in l.h.-side of Eq. (\ref{bsnew}) had a form of the Stieltjes transform which could be inverted analytically.
Inverting it, an equation, again in the form (\ref{L6}), was found. The operator, inverse to the operator in the l.h.-side of Eq. (\ref{bsnew}), applied to the  r.h.-side of (\ref{bsnew}), creates the kernel $N$ of the equation (\ref{L6}).  $N$ is expressed through $V$  as follows \cite{nak2}:
\begin{eqnarray} \label{L7}
&&N(\gamma,z,\gamma',z') =\frac{\gamma}{2\pi}\int_{-\pi+\epsilon}^{\pi-\epsilon} \;d\phi\;  e^{i\phi} 
\nonumber\\
&&\phantom{\frac{\gamma}{2\pi}\int_{-\pi}^{\pi} }\times V\Bigl(\gamma e^{i\phi}-z^2 m^2-(1-z^2)\kappa^2,z,\gamma',z'\Bigr)
\end{eqnarray}
with $\epsilon\to 0$.
The equations in the form (\ref{L6}), found in \cite{FrePRD14} and \cite{nak2}  by different methods, were equivalent since their kernels coincided with each other. 
However, these kernels were obtained in rather different forms, so it was uneasy to compare them analytically. Their equivalence was demonstrated in \cite{nak2} numerically.

For the first glance, the way from the kernel $K$ in the BS equation (\ref{bse}) to the kernel $N$ in the equation (\ref{L6}) was paved.
However, as it was noticed and demonstrated in examples already in
 \cite{nak2}, the direct calculation of the kernel $N(\gamma,z,\gamma',z')$ by Eq. (\ref{L7})  was, to some extent, ambiguous.  This problem arised since the kernel 
 $V(\gamma,z,\gamma',z')$  was expressed via the functions sqrt and $\log$, see Eqs. (\ref{W}) and (\ref{bb}) below, which are multivalued in the complex plane (where we go after the substitution $\gamma\to\gamma e^{i\phi}-z^2 m^2-(1-z^2)\kappa^2$). 
 Though this ambiguity can be eliminated, this delicate problem requires some extra efforts and very attentive and careful definitions of branches of the multivalued complex functions. 
 This complicates the direct practical applications of the approach \cite{nak2} to solving the BS equation via finding the function $g(\gamma,z)$ from Eq. (\ref{L6}).
 
The aim of the present work is to represent the kernel $N$ in Eq. (\ref{L6}) in another (third) form which does not contain this ambiguity. 
We emphasize that we are only talking about a new, more convenient \emph{form} of one and the same kernel.
The kernel $N$ in new form, derived in the present article, is identical to ones found in Refs. \cite{FrePRD14,nak2}. However, its form is completely unambiguous, restricted to the real functions only, does not require delicate choosing branches of multivalued functions and therefore it makes direct numerical solving Eq. (\ref{L6}) not more complicated than solving the Euclidean BS equation. Though the kernel $N$, derived below, might look a little bit lengthy and cumbersome, it is given by the explicit formulas, does not contain any integrations (for one-boson exchange) and therefore it is easy, unambiguously and rapidly computated.

The advantages of this new representation of the kernel $N$ in the equation (\ref{L6}) make solving the BS equation a simple and routine work.

Further content of the present paper is the following. In Sec. \ref{sec3} we derive this new form of the kernel $N$. Sec. \ref{summary} summarizes the results of Sec. \ref{sec3} in the form of recipes ready to use. In Sec. \ref{mu0} we take the limit of zero exchange mass $\mu\to 0$ and show analytically that in this limit the equation (\ref{L6}) with the kernel found in Secs. \ref{sec3} and \ref{summary} turns into the Wick-Cutkosky equation and kernel \cite{wick,cutk}. This serves as another test of our kernel  and shows that the equation (\ref{L6}) is a direct generalization of the Wick-Cutkosky one for massive exchange. 
In Sec. \ref{half} the equation (\ref{L6}) defined in the interval $-1\leq z\leq 1$ is reduced to the half interval $0\leq z\leq 1$ and the corresponding kernel is derived. This allows to increase the precision of  numerical calculations. In Sec. \ref{clk} we explain, how to derive the kernel in the canonical equation 
(\ref{L6}) from the Feynman cross-ladder one. The method can be applied to any kernel $K$ in the BS equation (\ref{bse}).  Sec. \ref{concl} contains the concluding remarks.

\section{Calculating the ladder kernel $N$  in the equation (\ref{L6})}\label{sec3}

The relation between the original kernel $K$, Eq. (\ref{K}),  appearing in the BS equation (\ref{bse}), and the  kernel $V$ in Eqs. (\ref{bsnew}), (\ref{L7})  was derived in Ref. \cite{bs1}.
In  the case of the ladder  kernel (\ref{K}), the kernel $V$ takes the form:
\begin{equation} \label{Kn}
V(\gamma,z,\gamma',z')=\left\{
\begin{array}{ll}
W(\gamma,z,\gamma',z'),&\mbox{if $-1\le z'\le z\le 1$}\\
W(\gamma,-z,\gamma',-  z'),&\mbox{if $-1\le z\le z'\le 1$}
\end{array}\right.
\end{equation}
where:
\begin{equation}\label{V1}
W(\gamma,z,\gamma',z')= \frac{\alpha m^2(1-z)^2} {2\pi [\gamma +z^2 m^2+(1-z^2)\kappa^2 ]} \int_0^1 dv \left(\frac{v}{D}\right)^2,
\end{equation}
\begin{eqnarray*}
D&=&v(1-v)(1-z')\gamma+v(1-z)\gamma'
\nonumber\\
&+&v(1-z)(1-z')\Bigl[1+z(1-v) +v z'\Bigr]\kappa^2  
\nonumber\\
 &+&v\Bigl[(1-v)(1-z')z^2+v{z'}^2(1-z)\Bigr]m^2
 \nonumber\\
 &+&(1-v)(1-z)\mu^2.
\end{eqnarray*}

\subsection{Integrating over $v,\phi$}\label{v_phi_2}
The integral  (\ref{V1}) over $v$ was calculated analytically \cite{bs1}. It reads:
\begin{widetext}
\small
\begin{equation}\label{W}
W(\gamma,z,\gamma',z') =
\frac{\alpha m^2}{2\pi}  \frac{(1-z)^2}{[\gamma+z^2m^2+(1-z^2)\kappa^2]} 
\frac{1}{b_2^2(b_+ -b_-)^3}
   \left[ \frac{(b_+ -b_-)(2b_+ b_- -b_+ -b_-)}{(1-b_+)(1-b_-)} \right. 
 +  \left.2b_+ b_- \log \frac{b_+ (1-b_-)}{b_- (1-b_+)}\right], 
\end{equation}
\end{widetext}
\normalsize
where
\begin{eqnarray}
b_0 &=& (1-z)\mu^2, 
\nonumber\\
 b_\pm &=& -\frac{1}{2b_2} \;\left( b_1 \pm \sqrt{b_1^2-4b_0b_2}\right),
\label{bb}\\
b_1 &=& \gamma+\gamma' - (1-z)\mu^2 - \gamma' z -\gamma z' 
\nonumber\\
&+&(1-z')\left[z^2m^2+(1-z^2)\kappa^2\right], 
\nonumber\\
b_2 &=& -\gamma (1-z')-(z-z') [  (1-z)(1-z')\kappa^2
\nonumber\\
&& \phantom{ z-\gamma (1)-(z-z')}
+(z+z'-zz') m^2].
\nonumber
\end{eqnarray}
The kernel $N(\gamma,z,\gamma',z')$ is now determined 
by the integral over $\phi$, Eq. (\ref{L7}).

As mentioned, Eqs. (\ref{W}) and (\ref{bb}) and, hence, the kernel $V(\gamma,z,\gamma',z')$ contains $\log$ and sqrt -- the multivalued functions in the complex plane.  This requires, after substituting 
$\gamma\to \gamma e^{i\phi}-z^2 m^2-(1-z^2)\kappa^2$, very careful definition of their branches. This is the reason of difficulty and limitation in practical use of this method. In the next section we will avoid this difficulty.

\subsection{Integrating over $\phi,v$}\label{phi_v_2}

The method  of calculation of the kernel $N$ used in the present work differs from one used in \cite{nak2} only by the order of integrations: first over $\phi$, then over $v$. This minimal, for the first glance, difference turns out to be crucial.  The integrand in Eq. (\ref{V1}), before integrating over $v$,  does not contain any multi-valued functions. Therefore the  $\phi$-integration is straightforward and
unambiguous. It results in real functions. The subsequent $v$-integration is analytical. This calculation of the double integral (\ref{L7}),  (\ref{V1}) will be carried out below.

 \bigskip

We take the kernel $V(\gamma,z,\gamma',z')$ in the form Eqs. (\protect{\ref{Kn}}), (\protect{\ref{V1}}), do not integrate over $v$ in  (\protect{\ref{V1}}), but make the substitution 
 $\gamma\to \gamma e^{i\phi}-z^2 m^2-(1-z^2)\kappa^2$ in the argument $\gamma$.  In this way, the kernel  (\protect{\ref{L7}}) obtains the form:
\begin{equation}\label{N0}
N(\gamma,z,\gamma',z') =\int_0^1n(\gamma, z,\gamma',z';v)\,dv,
\end{equation}
where
\begin{equation}\label{J}
n(\gamma,z,\gamma',z';v) =c(v)\int_{-\pi+\epsilon}^{\pi-\epsilon}d\phi\frac{1}{(a(v)\exp(i\phi)+b(v))^2}
\end{equation}
and 
\begin{eqnarray}\label{abc}
a(v)&=&\gamma v(1-v)(1-z'),
\nonumber\\
b(v)&=&(1-z)\Bigl[(1-v)\mu^2
\nonumber\\
&+&v\Bigl(\gamma'+v({z'}^2m^2+(1-{z'}^2)\kappa^2)\Bigr)\Bigr],
\nonumber\\
c(v)&=&\frac{1}{(2\pi)^2}\alpha m^2v^2(1-z)^2.
\end{eqnarray}
Note that  $a(v)$, $b(v)$ and $c(v)$ are positive. {Besides the argument $v$,  $a(v)$ depends also on 
$\gamma,z'$, $b(v)$ depends on $z,\gamma',z'$ and $c(v)$ depends on $z$. This creates dependence of  $N(\gamma,z,\gamma',z')$ on all four variables 
$\gamma,z,\gamma',z'$.
 We omit these extra arguments in $a(v),b(v),c(v)$ for shortness.  For some values of parameters and variables it may happen that $a(v)=b(v)$. Then the denominator in (\ref{J}) crosses zero when $\phi\to \pm \pi$ ($\epsilon \to 0$) and the integrand of $n(\gamma,z,\gamma',z';v)$ vs. $\phi$ becomes singular. It remains to be singular after $\phi$-integration. We will show that the singularity, which survives after the $\phi$-integration,  is represented by the delta-function $\delta[b(v)-a(v)]$ and after integration over $v$ in (\ref{N0}) 
it gives a finite, non-singular contribution to $N(\gamma,z,\gamma',z')$. However, due to its origin,  we will call it "singular part contribution".  Another finite contribution results from the non-singular part of $n(\gamma,z,\gamma',z';v)$ vs. $v$. So, we get two contributions, 
$N_{reg}$ and $N_{sing}$, which we will call non-singular (or regular) and singular part contributions correspondingly.
The kernel $N(\gamma,z,\gamma',z')$ is the sum of these two terms:
\begin{equation}\label{N1}
N(\gamma,z,\gamma',z') =N_{reg}+N_{sing},
\end{equation}
where 
\begin{eqnarray}\label{N1a} 
N_{reg}&=&\int_{v_1}^{v_2} n_{reg}(\gamma,z,\gamma',z';v) \,dv,
\\
N_{sing}&=&\int_0^1 n_{sing}(\gamma,z,\gamma',z';v)\,dv.
\label{Ns}
\end{eqnarray}
The functions $n_{reg}(\gamma,z,\gamma',z';v)$ and  $n_{sing}(\gamma,z,\gamma',z';v)$ will be found below and are given by eqs. (\protect{\ref{Jv})} and (\protect{\ref{j2f}}) correspondingly. The arguments of the functions $N_{reg}$ and $N_{sing}$ are omitted here. 
We precise and  restore them below, calculating these functions. Note that, depending on the argument values, the contribution of $N_{sing}$ may be zero. 
Or, on the contrary, $N_{reg}$ and $N_{sing}$ may enter in $N(\gamma,z,\gamma',z')$ twice, for different sets of arguments (see Sec. \ref{summary}).
In the integral (\ref{N1a}) for $N_{reg}$ the integration interval $0\leq v\leq 1$ was replaced by the interval $0\leq v_1\leq v \leq v_2\leq 1$ since, as it will be shown below, 
$n_{reg}(\gamma,z,\gamma',z';v)$ contains the theta-function which can make the integration interval more narrow than $0\leq v\leq 1$ and reduce the integration limits (though, not always). 

We start with calculation of the non-singular part $N_{reg}$. Calculating it, we will also determine the integration limits $v_{1,2}$.
\bigskip

\centerline{\it Non-singular part contribution}
 \bigskip

To calculate the integral (\ref{J}) for $a(v)\neq b(v)$, we can put in (\ref{J}) $\epsilon=0$.
 We introduce the complex variable $y=\exp(i\phi)$. Then this integral obtains the form:
\begin{eqnarray}\label{J1}
n_{reg}(\gamma,z,\gamma',z';v)&=&\frac{1}{i}\int_{\mathcal C}dy\frac{c(v)}{y[a(v)y+b(v)]^2},
\end{eqnarray}
where integration is carried out over the unit closed circle ${\mathcal C}$ around origin  in the complex plane $y$. The integrand vs. $y$ has two poles:
$$ 
y=0\quad \mbox{and} \quad y=-\frac{b(v)}{a(v)}
$$
with the corresponding residues:
$$
Res(y=0)=-\frac{ic(v)}{b^2(v)} \quad \mbox{and}\quad Res\left(y=-\frac{b(v)}{a(v)}\right)=
\frac{ic(v)}{b^2(v)}
$$
If both poles are within the unit circle ${\mathcal C}$, the residues cancel each other and  the integral is zero. The result is not zero if the pole $y=-\frac{b(v)}{a(v)}$ is outside
 the unite circle ${\cal C}$. It happens when $-\frac{b(v)}{a(v)}<-1 \to b(v)>a(v)$.
Therefore, the result is:
\begin{eqnarray}\label{Jv}
n_{reg}(\gamma,z,\gamma',z';v)&=&\left.2\pi i Res(y=0)\right|_{b(v)>a(v)}
\nonumber\\
&=& \displaystyle{\frac{2\pi c(v)}{b^2(v)}}\theta[b(v)-a(v)].
\end{eqnarray}
Its contribution to the kernel $N(\gamma,z,\gamma',z')$ is given by the integral (\ref{N1a})  over $v$. This integral can be calculated analytically, 
in terms of the primitives, since $b(v)$ vs. $v$  is a quadratic polynomial. 

Some tedious analysis is related to the theta-function in (\ref{Jv}). Its argument $b(v)-a(v)$ is also a quadratic function of $v$:
\begin{equation}\label{equ}
b(v)-a(v)=A_0 v^2+B_0 v+C_0=A_0(v-v_-)(v-v_+),
\end{equation}
where 
\begin{eqnarray}
A_0&=& \gamma(1-z')+(1-z)[m^2{z'}^2+\kappa^2(1-{z'}^2)] >0,\;\;
\label{A0}
\\
B_0&=&(\gamma'-\mu^2)(1-z)-\gamma(1-z'),\;
\label{B0}
\\
C_0&=&\mu^2(1-z)>0
\label{C0}
\end{eqnarray}
and $v_{\mp}$ are the roots of the quadratic equation \mbox{$b(v)-a(v)=0$}:
\begin{equation}
v_{\mp}=\frac{1}{2A_0}(-B_0\mp\sqrt{D_0}),
\label{roots}
\end{equation}
where
\begin{equation}
D_0=B_0^2-4A_0C_0.
\label{D0}
\end{equation}
If $D_0<0$, the roots are complex and the polynomial (\ref{equ}) vs. $v$ does not change the sign. 
Since $A_0>0$,  it is always positive. In this case, the theta-function in (\ref{Jv}) does not give any constraint and the integration for $N_{reg}$ 
in (\ref{N1a})  is carried out in the interval $0\leq v\leq 1$, that is $v_1=0,v_2=1$.

If $D_0>0$, the equation 
$b(v)-a(v)=0$ vs. $v$  has two real roots $v_{\mp}$.  These roots, depending on values of the variables 
$\gamma,z,\gamma',z'$, can be inside the interval $0\leq v\leq 1$, one inside, other outside, etc. 
When $v$ crosses a root, the argument $b(v)-a(v)$ of the theta-function  changes the sign and the theta-function changes its value from 0 to 1 or back. Since the coefficient $A_0$ at $v^2$ in Eq. (\ref{equ}), is positive, the value of polynomial between the roots $v_{\mp}$ is negative. If the interval $v_-\leq v\leq v_+$ 
is inside of $0\leq v\leq 1$ or  partially overlaps with it, their overlapping part (where $\theta[b(v)-a(v)]=0$) is excluded from integration.
In this way, the initial integration interval $0\leq v\leq 1$ is reduced by the theta-function to one or two smaller intervals.
\bigskip

Note that according to Eq. (\ref{roots}), if $D_0>0$ and since the product $A_0 C_0$ is positive,  then \mbox{$\sqrt{D_0}=\sqrt{B_0^2-4A_0C_0}<|B_0|$} and therefore
 the roots $v_{\mp}$ cannot have opposite signs: either both roots are negative (if $B_0>0$), or both are positive (if $B_0<0$). Therefore we will consider the following cases only:
\begin{enumerate}
\item \label{case1}
$v_-<\;v_+<0$. 
Since the interval between the roots $v_-<v<v_+$ (where $b(v)-a(v)<0 \to \theta[b(v)-a(v)]=0$) does not overlap with $0\leq v\leq 1$, whereas 
for $v>v_+$ (and certainly, for $v>0$), $b(v)-a(v)>0 \to \theta[b(v)-a(v)]=1$, in this case, the theta-function in (\ref{Jv}) does not give any constraint and the integration limits in (\ref{N1a}) are $v_1=0$, $v_2=1$.
\item \label{case4}
$0<v_- <v_+<1$. The theta-function is zero if $v_-<v<v_+$. The integration limits are $v_1=0$, $v_2=v_-$ and $v_1=v_+$, $v_2=1$. One should take sum of the integrals over these two intervals.
 \item \label{case5}
$0<v_-<1<v_+$. The integration limits are $v_1=0$, $v_2=v_-$.
\item \label{case6}
$1<v_-<v_+$.  The integration limits are $v_1=0$, $v_2= 1$, like in the case \ref{case1}.
\end{enumerate}

The integral (\protect{\ref{N1a}}) with $n_{reg}$ defined in (\ref{Jv}) is represented as:
\begin{equation}\label{Nreg}
N_{reg}(\gamma,z,\gamma',z';v_1,v_2) 
=F(v_2)-F(v_1),
\end{equation}
where $F(v)$ is the primitive:
\begin{equation}\label{Fv}
F(v)=\int\frac{2\pi c(v)dv}{b^2(v)}=
\frac{\alpha m^2}{2\pi}\int\frac{v^2 dv}{(A_1v^2+B_1v+C_1)^2}
\end{equation}
and $v_{1,2}$ are the integration limits indicated above for the cases \ref{case1}-\ref{case6}.
The denominator in r.h.-side of (\ref{Fv}) contains
\begin{eqnarray}
\frac{b(v)}{1-z}&=&(1-v)\mu^2+v\Bigl[\gamma'+v\Bigl({z'}^2m^2+(1-{z'}^2)\kappa^2\Bigr)\Bigr]
\nonumber\\
&=&A_1v^2+B_1v+C_1,
\label{ABC1}
\end{eqnarray}
where
\begin{eqnarray*}
&&A_1=m^2{z'}^2+\kappa^2(1-{z'}^2)>0,
\nonumber\\
&&B_1=\gamma'-\mu^2,
\nonumber\\
&&C_1=\mu^2>0.
\end{eqnarray*}
From the expression (\ref{ABC1}) 
we see that since $\gamma'>0$, in the interval $0\leq v\leq 1$ (which is larger than the integration interval $v_1\leq v\leq v_2$ in (\ref{N1a})), 
the value $A_1v^2+B_1v+C_1$ is always positive.  This means that if the denominator in (\ref{Fv})  has zeros,  these zeros are outside the interval \mbox{$0\leq v\leq 1$}.  Therefore, the integrand in (\ref{Fv}) is not singular. However, the form of $F(v)$ depends on the sign of 
\begin{equation}
D_1=B_1^2-4A_1C_1.
\label{D1}
\end{equation}
Namely, calculating the integral (\ref{Fv}), we find:
\begin{equation}\label{Fv2}
F(v)=
\left\{
\begin{array}{ll}
F^-(v),& \mbox{if $D_1<0$}
\\
F^+(v),& \mbox{if $D_1>0$}
\end{array}\right.,
\end{equation}

where
\begin{eqnarray}
F^-(v)&=&\frac{\alpha m^2}{2\pi}\left[\frac{B_1 C_1+(B_1^2-2A_1C_1)v}{A_1|D_1|[C_1+v(B_1+vA_1)]}\right.
\nonumber\\
&+&\left.\frac{4C_1}{|D_1|^{3/2}}\arctan\frac{B_1+2vA_1}{\sqrt{|D_1|}}\right]
\label{Fvm}
\end{eqnarray}
and
\begin{eqnarray}
F^+(v)&=&-\frac{\alpha m^2}{2\pi}\left[\frac{B_1C_1+(B1^2-2A_1C_1)v}{A_1D_1[C_1+v(B_1+vA_1)]}\right.
\nonumber\\
&+&\left.\frac{2C_1}{D_1^{3/2}}\log\left(1-\frac{2\sqrt{D_1}}{B_1+\sqrt{D_1}+2A_1 v}\right)\right].
\label{Fvp}
\end{eqnarray}

The equation (\ref{Nreg}) together with Eqs.  (\ref{Fv2}-\ref{Fvp}) determine the regular part contribution $N_{reg}$.
\bigskip

\centerline{\it Singular part contribution}
\bigskip

The singular part contribution results from the integral (\ref{J}) where $b(v)= a(v)$ for some value of $v$. To extract it from (\ref{J}), we calculate this integral analytically and then take the limit 
$b(v)\to a(v)$. Calculating the integral via primitive, we find:
\begin{widetext}
\begin{eqnarray}
n(\gamma,z,\gamma',z';v)&=&\frac{2c(v)(\pi-\epsilon)}{b^2(v)}+\frac{c(v)Arg\Bigl[b+a(v)\exp[i(\epsilon-\pi)]\Bigr]}{b^2(v)}-\frac{c(v)Arg\Big[(b(v)+a(v)\exp[i(-\epsilon+\pi)]\Bigr]}{b^2(v)}
\nonumber\\
&-&\frac{2a(v)c(v)\sin\epsilon}{b(v)[(b(v)-a(v)\cos\epsilon)^2+a^2(v)\sin^2\epsilon]}.
\label{nv}
\end{eqnarray}
\end{widetext}
At $\epsilon\to 0$ and $b(v)\to a(v)$, the last term in (\ref{nv}) only is singular and dominates.  We keep it, take the limit $\epsilon\to 0$ and make the replacement $a(v)\epsilon=\epsilon'$.
In this way we find:
$$
n_{sing}(\gamma,z,\gamma',z';v)=-\frac{2c(v)\epsilon'}{b(v)[(b(v)-a(v))^2+{\epsilon'}^2]}.
$$
Using the formula: 
$$
\frac{1}{\pi}\frac{\epsilon}{(x^2+\epsilon^2)}\stackrel{\epsilon\to 0}{=}\delta(x),
$$
we obtain
\begin{equation}\label{j2f}
n_{sing}(\gamma,z,\gamma',z';v)=-\frac{2\pi c(v)}{a(v)}\delta[b(v)-a(v)].
\end{equation}
The difference $b(v)-a(v)$ (a second order polynomial in $v$) is represented in the form of product,  Eq. (\ref{equ}).
If roots $v_{\mp}$ of the equation $b(v)-a(v)=0$ vs. $v$ are real and both are in the interval $0<v<1$, then  the integral over $v$ (the second term in (\ref{N1a})) is reduced to the sum over roots:
\small
\begin{eqnarray}\label{j22a}
\int_0^1n_{sing}(\gamma,z,\gamma',z';v)dv&=&N_{sing}(\gamma,z,\gamma',z';v_-)
\nonumber\\
&+&N_{sing}(\gamma,z,\gamma',z';v_+)
\end{eqnarray}
where we denoted
\begin{equation}\label{Nsing}
N_{sing}(\gamma,z,\gamma',z';v)=
-\frac{2\pi c(v)}{A_0|v_+ -v_-|\,a(v)},
\end{equation}
\normalsize
$a(v),\,c(v)$ are defined in (\ref{abc}), $A_0$ is defined in (\ref{A0}).
If only one root $v_{-}$ is in the interval $0<v<1$, then instead of the sum (\ref{j22a}) we should take the contribution of this root only:
\begin{equation}\label{j22b}
\int_0^1n_{sing}(\gamma,z,\gamma',z';v)dv=N_{sing}(\gamma,z,\gamma',z';v_-).
\end{equation}
As explained above, two real roots $v_{\mp}$ have the same sign. Therefore the situation when one root $v_+$ is in the interval $0<v<1$, whereas the root $v_-$ is outside (and, hence, $v_-<0$) is impossible. 

 If both roots are outside of the interval $0<v<1$, the singular part does not contribute. This analysis is similar to one given in the points
\ref{case1}-\ref{case6} above Eq.   (\ref{Nreg}).

In this way, the equations  \mbox{(\protect{\ref{j22a}}-\protect{\ref{j22b}})} determine the singular part contribution $N_{sing}$.

\section{Summary}\label{summary}
In this section, summarizing the above results,  we give the recipe of calculating the ladder massive exchange kernel 
$N(\gamma,z,\gamma',z')$ in the equation (\ref{L6}). It takes a few following steps. Depending on the relation between $z,z'$ ($z<z'$ or $z>z'$), the kernel $N(\gamma,z,\gamma',z')$ is expressed, by the formula similar to eq. (\ref{Kn}),  via the auxiliary kernel $\tilde{N}(\gamma,z,\gamma',z')$ which will be constructed in this section. In its turn, the kernel $\tilde{N}(\gamma,z,\gamma',z')$ is expressed via $N_{reg}(\gamma,z,\gamma',z';v_1,v_2)$, eq. (\ref{Nreg}), and via 
$N_{sing}(\gamma,z,\gamma',z';v)$, eq. (\ref{Nsing}). The function $N_{reg}(\gamma,z,\gamma',z';v_1,v_2)$ is constructed via another function $F(v)$,  eq. (\ref{Fv2}). The contributions  $N_{reg}$ and $N_{sing}$ to $\tilde{N}$  depend on the sign of $D_0$, Eq. (\ref{D0}), determining the existence of the real roots 
$v_{\mp}$ of the equation $b(v)-a(v)=0$. If $D_0>0$ and the real roots exist, the result depends on their positions relative to the interval $0\leq v \leq 1$. For all these cases the kernel $N(\gamma,z,\gamma',z')$ in eq. (\ref{L6}) is defined below. The construction is a little bit lengthy, but it is coherent and absolutely unambiguous.

We do not stop at the contruction of the kernel $N(\gamma,z,\gamma',z')$, but we reduce below the equation  (\ref{L6}) to the equation (\ref{eq6}) defined in the half-interval $0\leq z \leq 1$.  The kernel $N_{half}(\gamma,z,\gamma',z')$ in this reduced equation will be also expressed through $\tilde{N}(\gamma,z,\gamma',z')$.
 
We consider the following cases.
\begin{enumerate}
\item
\label{item1}
$D_0<0$. According to eq. (\ref{roots}), the roots $v_{\mp}$ are complex.
Therefore, the singular contribution $N_{sing}$ is absent. Besides, $b(v)-a(v)>0$. Hence, always
$\theta[b(v)-a(v)]=1$.
Therefore
\begin{equation}\label{Nt1}
\tilde{N}(\gamma,z,\gamma',z')= N_{reg}(\gamma,z,\gamma',z';v_1=0,v_2=1)
\end{equation}
with $N_{reg}(\gamma,z,\gamma',z';v_1,v_2)$ defined in (\ref{Nreg}).
\item
\label{item2}
$D_0>0$.
 The roots $v_{\mp}$ given by eq. (\ref{roots}) are real. The form of the kernel $\tilde{N}$ is different in three following sub-cases, depending on the positions of the roots.
 
(a) The roots are  out of the integration domain \mbox{$0\leq v\leq 1$}  over $v$: either  \mbox{$v_-<v_+<0$}, or 
 \mbox{$1<v_- <v_+$}.
Then the kernel $\tilde{N}(\gamma,z,\gamma',z')$ is still defined by Eq. (\ref{Nt1}).

(b) Both roots are in the integration interval:
 \mbox{$0<v_-<v_+<1$}. Then the kernel $\tilde{N}(\gamma,z,\gamma',z')$ reads
\begin{widetext}
\begin{eqnarray}\label{Nt2}
\tilde{N}(\gamma,z,\gamma',z')&=& N_{reg}(\gamma,z,\gamma',z';v_1=0,v_2=v_-)+N_{reg}(\gamma,z,\gamma',z';v_1=v_+,v_2=1)
\nonumber\\
&+&N_{sing}(\gamma,z,\gamma',z';v_-)+N_{sing}(\gamma,z,\gamma',z';v_+).
\end{eqnarray}
\end{widetext}
with $N_{reg}(\gamma,z,\gamma',z';v_1,v_2)$ defined in (\protect{\ref{Nreg}}) and $N_{sing}(\gamma,z,\gamma',z';v)$ defined in (\protect{\ref{Nsing}}).

(c) One root is in the integration interval  \mbox{$0\leq v\leq 1$}, whereas the second one is at $v>1$. That is:
 \mbox{$0<v_-<1<v_+$}.
Then the kernel $\tilde{N}(\gamma,z,\gamma',z')$ takes the form
\begin{eqnarray}\label{Nt3}
\tilde{N}(\gamma,z,\gamma',z')&=& N_{reg}(\gamma,z,\gamma',z';0,v_-)
\nonumber\\
&+&N_{sing}(\gamma,z,\gamma',z';v_-).
\end{eqnarray}
\end{enumerate}
We remind that the roots cannot have opposite signs (see remark below eq. (\ref{D0})), therefore the case $v_-<0<v_+$ is excluded.
\bigskip

The kernel $N(\gamma,z,\gamma',z')$ is constructed via $\tilde{N}(\gamma,z,\gamma',z')$ similarly to eq. (\ref{Kn}):
\begin{equation} \label{Knew}
N(\gamma,z,\gamma',z')=\left\{
\begin{array}{ll}
\tilde{N}(\gamma,z,\gamma',z'),&\mbox{if $-1\le z'\le z\le 1$}\\
\tilde{N}(\gamma,-z,\gamma',-  z'),&\mbox{if $-1\le z\le z'\le 1$}
\end{array}\right.
\end{equation}
We emphasize that in the second case ($\mbox{if $-1\le z\le z'\le 1$}$) we should make the replacement $z\to -z,\;z'\to -z'$ everywhere, including Eqs. (\ref{A0}-\ref{C0}) determining the positions of the roots (\ref{roots}).

This completes the calculations. Just this kernel $N(\gamma,z,\gamma',z')$, with $\tilde{N}(\gamma,z,\gamma',z')$ defined above, enters in the equation  (\ref{L6}) for 
$g(\gamma,z)$.
\bigskip

{\it Test 1.}
For test, we take the same parameters which were used in the test carried out   in \cite{nak2}, namely: 
$$
\alpha = 1,  m = 1, \mu = 0.15, M=1.9.
$$ 
We also take the same values of variables:
$$
 \gamma = 0.1, \gamma' = 1, z = 0.2, z' = 0.35.
 $$
Since $z<z'$,  according to Eq. (\ref{Knew}), the kernel $N(\gamma,z,\gamma',z')$  is determined by $\tilde{N}(\gamma,-z,\gamma',-z')$.
I.e. in further calculations,  we should change the signs of the variables $z,z'$: $z\to -z$, $z'\to -z'$.
For these new  values of variables, we find $D_0=1.036>0$.
By Eq. (\ref{roots}), we find the roots of the equation $b(v)-a(v)=0$: $v_-=-2.672$, $v_+=-0.0263$. Hence, we deal
with the case \ref{item2}(a).  The value of $\tilde{N}$ is determined through $N_{reg}$ by Eq. (\ref{Nt1}).
The contribution  $N_{sing}$ of the singular part is absent.
 In its turn, $N_{reg}$
is determined by Eqs. (\ref{Nreg}), (\ref{Fv2}) and, since $D_1=0.937>0$,   finally  by $F_+(v)$, Eq. (\ref{Fvp}).  In this way, we find the value of the kernel: $N=0.1153980591510$, that coincides within all digits with the value found in Eq. (31) of Ref. \cite{nak2}. 

{\it Test 2.} For the same set of the parameters  $\alpha,m,\mu,M$ we take the following set of variables: 
$$
 \gamma = 0.5, \gamma' = 0.1, z = 0.2, z' = 0.35.
$$
Since $z<z'$,  according to Eq. (\ref{Knew}), the kernel $N(\gamma,z,\gamma',z')$  is still determined by $\tilde{N}(\gamma,-z,\gamma',-z')$.
I.e. in further calculations,  we should take the negative values  of $z,z'$: $z\to -z$, $z'\to -z'$.
For these new  values of variables, we find $D_0=0.239>0$.
By Eq. (\ref{roots}), we find the roots of the equation $b(v)-a(v)=0$: $v_-=0.050$, $v_+=0.579$.
Hence, we now deal with the case \ref{item2}(b). The value of $\tilde{N}$ is determined through $N_{reg}$ and $N_{sing}$
by Eq. (\ref{Nt2}). The singular part now gives a non-zero contribution. 
 In its turn, $N_{reg}$
is determined by Eqs. (\ref{Nreg}), (\ref{Fv2}) and, since $D_1=-0.0127<0$,  finally by $F_-(v)$, Eq. (\ref{Fvm}). 
In this way, we find the value of the kernel: $N=-0.0531890858160$, that coincides within all digits with the value found in Eq. (33) of Ref. \cite{nak2}. 

Note that these two tests cover all three functions used to calculate $N$: the function  $F^-(v)$, Eq. (\ref{Fvm}),  the function  $F^+(v)$, Eq. (\ref{Fvp}), 
and $N_{sing}(\gamma.z,\gamma',z',v)$, Eq. (\ref{Nsing}).

\section{Massless exchange}\label{mu0}
The case $\mu=0$ corresponds to the Wick-Cutkosky model \cite{wick,cutk}. For simplicity, we will consider the ground state only. In this model, the ground state function $g(\gamma,z)$ turns into 
\begin{equation}\label{g0}
g(\gamma,z)=\delta(\gamma)g(z)
\end{equation}
and the equation for $g(z)$ was found in 
\cite{wick,cutk}. Below we will show that the equation (\ref{L6}) with the kernel found in the previous sections turns, in the limit $\mu\to 0$, into the Wick-Cutkosky equation.

According to Sec. \ref{summary}, the formulas which should be used to calculate  the kernel $N$ in the equation (\ref{L6}) depend on the position of the roots, Eq. (\ref{roots}), relative to the interval $0<v<1$. Decomposing $v_\mp$, Eq. (\ref{roots}), in series of $\mu^2$ and denoting $B_{00}=B_0(\mu=0)$,
we should consider two cases: $B_{00}<0$ and $B_{00}>0$. 
\bigskip

The case
$B_{00}<0$  $\left(\gamma'<\frac{1-z'}{1-z}\gamma\right)$:
$$
 v_- =-\frac{1}{B_{00}}(1-z)\mu^2>0,\;v_+ = -\frac{B_{00}}{A_0}>0.
$$
From Eqs. (\ref{A0}), (\ref{B0}) it follows: 
$
A_0>0,\;A_0+B_{00}>0. 
$
Therefore
 $-\frac{B_{00}}{A_0}<1$,  $0<v_-<v_+<1$. We should calculate the kernel according to the case 2(b) from the Summary, Sec. \ref{summary}, i.e., by Eq. (\ref{Nt2}).
 \bigskip

The case
$B_{00}>0$ $\left(\gamma'>\frac{1-z'}{1-z}\gamma\right)$:
$$
v_- = -\frac{B_{00}}{A_0}<0,\;v_+ = -\frac{1}{B_{00}}(1-z)\mu^2<0.
$$
In this case: $v_-<v_+<0$. We should calculate the kernel according to the case 2(a) from  Sec. \ref{summary}, 
i.e., by Eq. (\ref{Nt1}).
\bigskip

In this way, taking the limit $\mu\to 0$ in Eq. (\ref{Nt1}) and in the first line of Eq. (\ref{Nt2}), we find the regular part contribution:
\begin{widetext}
\begin{equation}\label{N12}
N_{reg}(\gamma,z,\gamma',z') =
\left\{
\begin{array}{ll}
\frac{\alpha m^2(1-z)}{2\pi\gamma(1-z')}\frac{1}{\gamma'+m^2{z'}^2+\kappa^2(1-{z'}^2)},&\mbox{if  $\gamma'<\frac{1-z'}{1-z}\gamma$}
\\
\frac{\alpha m^2}{2\pi\gamma'}\frac{1}{\gamma'+m^2{z'}^2+\kappa^2(1-{z'}^2)},&\mbox{if  $\gamma'>\frac{1-z'}{1-z}\gamma$}
\end{array}
\right.
\end{equation}
\end{widetext}

The singular part contribution is calculated similarly.
For $B_0<0$ $\left(\gamma'<\frac{1-z'}{1-z}\gamma\right)$ this contribution is given by the second line of eq.~(\ref{Nt2}).
For $B_0>0$ $\left(\gamma'>\frac{1-z'}{1-z}\gamma\right)$ it  is absent, see eq.~(\ref{Nt1}).
 That is:
\begin{widetext}
\begin{equation}\label{Ns_1}
N_{sing}(\gamma,z,\gamma',z') =
\left\{
\begin{array}{ll}
-\frac{\alpha m^2(1-z)}{2\pi\gamma(1-z')}\frac{1}{\gamma'+m^2{z'}^2+\kappa^2(1-{z'}^2)},&\mbox{if  $\gamma'<\frac{1-z'}{1-z}\gamma$}
\\
0&\mbox{if  $\gamma'>\frac{1-z'}{1-z}\gamma$}
\end{array}
\right.
\end{equation}
\end{widetext}
The full kernel $\tilde{N}$ is the sum of (\ref{N12}) and (\ref{Ns_1}):
 \begin{eqnarray}\label{N_1}
\tilde{N}(\gamma,z,\gamma',z') &=&
\left\{
\begin{array}{ll}
0,&\mbox{\small if  $\gamma'<\frac{1-z'}{1-z}\gamma$}
\\
\frac{\alpha m^2}{2\pi\gamma'}\frac{1}{\gamma'+m^2{z'}^2+\kappa^2(1-{z'}^2)},&\mbox{\small if  $\gamma'>\frac{1-z'}{1-z}\gamma$}
\end{array}
\right.
\nonumber\\
&=&\frac{\alpha m^2}{2\pi\gamma'}\;\frac{\theta[\gamma'(1-z)-\gamma(1-z')]}{\gamma'+m^2{z'}^2+\kappa^2(1-{z'}^2)}
\end{eqnarray}
The first lines in Eqs. (\ref{N12}) and (\ref{Ns_1}) are the same, up to the opposite signs,  they cancel each other in Eq. (\ref{N_1}).

We will show that the solution in the Wick-Cutkosky form (\ref{g0}) indeed satisfies the equation (\ref{L6}) with the  kernel (\ref{N_1}).
Since the support of the function $g(\gamma,z)$ is $\gamma>0$, the delta-function $\delta(\gamma)$ is obtained in the limit $\mu\to 0$ from a function having a peak at $\gamma>0$. 
To keep this property, we replace $\delta(\gamma)$ by  $\delta(\gamma-\epsilon)$ with $\epsilon>0$ and at the end of calculation take the limit $\epsilon\to 0$.
With the kernel $\tilde{N}(\gamma,z,\gamma',z')$, Eq. (\ref{N_1}), and with  $g(\gamma,z)=\delta(\gamma-\epsilon)g(z)$ the equation (\ref{L6}), after integration in r.h.-side over $\gamma'$ by means of the delta-function $\delta(\gamma'-\epsilon)$,  obtains the form:
$$
\delta(\gamma-\epsilon)g(z)=\int_{-1}^1dz' \frac{\alpha m^2}{2\pi\epsilon}\;\frac{\theta[\epsilon(1-z)-\gamma(1-z')]}{\epsilon+m^2Q(z')}g(z'),
$$
where $Q(z')=1-\eta^2(1-{z'}^2)$, $\eta=\frac{M}{2m}$.
Note that the function 
$$
h(\gamma)=\frac{1}{\epsilon}\theta[\epsilon(1-z)-\gamma(1-z')]
$$
vs. $\gamma$ differs from zero in the narrow interval $0<\gamma<\epsilon\frac{1-z}{1-z'}$. In this interval it is constant (equal to $\frac{1}{\epsilon}$) and it is zero outside.  The integral:
$$
\int_{-\infty}^{\infty}h(\gamma)d\gamma=\int_{0}^{\epsilon\frac{1-z}{1-z'}}h(\gamma)d\gamma=\frac{1-z}{1-z'}
$$
Hence, the integral over $\gamma$ from the function $\frac{1-z'}{1-z}h(\gamma)$ is 1, this function approximates the delta-function $\delta(\gamma-\gamma_0)$, where $\gamma_0$ is a value within the interval $0<\gamma_0< \epsilon\frac{1-z}{1-z'}$.
Therefore, replacing,  in the limit $\epsilon\to 0$, $h(\gamma)\to\frac{1-z}{1-z'}\delta(\gamma)$ and omitting 
$\delta(\gamma)$ in both parts of equation, we obtain the following equation for $g(z)$:
$$
g(z)=\frac{\alpha}{2\pi}\int_{-1}^1dz'\;\frac{1-z}{1-z'}\;\frac{g(z')}{Q(z')}.
$$

This equation is valid when $z'<z$. For $z'>z$ one should make the replacement $z\to -z$, $z'\to -z'$. 
Therefore we obtain:
\begin{equation}\label{WC}
g(z)=\frac{\alpha}{2\pi}\int_{-1}^1dz'\;R(z,z')\frac{g(z')}{Q(z')},
\end{equation}
where 
$$
R(z,z')=
\left\{
\begin{array}{ll}
\frac{1-z}{1-z'},& \mbox{if $z'<z$}
\\
\frac{1+z}{1+z'},& \mbox{if $z'>z$}
\end{array}
\right.
$$
The equation (\ref{WC}) exactly coincides with Eq. (15) from \cite{cutk}, derived in the case $\mu=0$,  and with Eq. (12) from \cite{cks_epjc} (for the quantum number $n=1$).
This coincidence can be considered as another test of the kernel presented in  Sec. \ref{summary}. 

\section{Equation on the half-interval ${\bf 0\le z\leq 1}$}\label{half}

It is easy to show that the solution of Eq. (\ref{L6}) with the kernel constructed by Eq. (\ref{Knew}), even with arbitrary function $\tilde{N}$, is either symmetric, or antisymmetric. For finding its solution numerically, it is useful, using this symmetry, to reduce it, instead of the interval \mbox{$-1\leq z\leq 1$}, to the half-interval $0 \leq z\leq 1$.

Using the definition (\ref{Knew}), we rewrite Eq. (\ref{L6}), in terms of the kernel $\tilde{N}$ as follows:
\begin{widetext}
\begin{eqnarray} \label{eq5}
g(\gamma,z)&=&\int_0^{\infty}d\gamma'\int_{-1}^{0}dz'\;\tilde{N}(\gamma,z,\gamma',z') g(\gamma',z')
\\
&+&\int_0^{\infty}d\gamma'\int_{0}^{z}dz'\;\tilde{N}(\gamma,z,\gamma',z') g(\gamma',z')
+ \int_0^{\infty}d\gamma'\int_{z}^{1}dz'\;\tilde{N}(\gamma,-z,\gamma',-z') g(\gamma',z')
\nonumber
\end{eqnarray}
\end{widetext}

For symmetric $g_{sym}(\gamma,z)=g_{sym}(\gamma,-z)$, making the replacement of variable $z'\to -z'$ in the first line of r.h.-side of Eq.  (\ref{eq5}), we get:
\begin{widetext}
\begin{eqnarray} \label{eq5a}
g_{sym}(\gamma,z)&=&\int_0^{\infty}d\gamma'\int_{0}^{1}dz'\;\tilde{N}(\gamma,z,\gamma',-z') g_{sym}(\gamma',z')
\\
&+&\int_0^{\infty}d\gamma'\int_{0}^{z}dz'\;\tilde{N}(\gamma,z,\gamma',z') g_{sym}(\gamma',z')
+ \int_0^{\infty}d\gamma'\int_{z}^{1}dz'\;\tilde{N}(\gamma,-z,\gamma',-z') g_{sym}(\gamma',z').
\nonumber
\end{eqnarray}
Introducing the kernel defined on the half-intervals \mbox{$0\leq z\leq 1$},  \mbox{$0\leq z'\leq 1$}
\begin{equation}\label{eq7}
N_{half}(\gamma,z,\gamma',z')=
\left\{\begin{array}{ll}
\tilde{N}(\gamma,z,\gamma',-z') +\tilde{N}(\gamma,z,\gamma',z'),& \mbox{if $z'<z$}
\\
\tilde{N}(\gamma,z,\gamma',-z') +\tilde{N}(\gamma,-z,\gamma',-z'),& \mbox{if $z'>z$}
\end{array}
\right.,
\end{equation}
\end{widetext}
where $\tilde{N}$ is defined in Eqs. (\ref{Nt1}), (\ref{Nt2}) or (\ref{Nt3}), depending on the positions of the roots $v_{\mp}$
(for given arguments  $\gamma,\pm z,\gamma',\pm z'$),
we rewrite the equation (\ref{eq5a}) as
\begin{equation} \label{eq6}
g_{sym}(\gamma,z)=\int_0^{\infty}d\gamma'\int_{0}^{1}dz'\;N_{half}(\gamma,z,\gamma',z') g_{sym}(\gamma',z').
\end{equation}

The antisymmetric solution $g_{asym}(\gamma,z)$ is determined by the equation  (\ref{eq6}) with the kernel $N_{half}$ which differs from (\ref{eq7})
by the opposite sign at the first term in r.h.-side in (\ref{eq7}), i.e., by the replacement $\tilde{N}(\gamma,z,\gamma',-z') \to -\tilde{N}(\gamma,z,\gamma',-z')$.

The equation (\ref{eq6}) is most convenient for the numerical solution, since its  interval in $z$ is twice more narrow than in (\ref{L6}). That allows to twice increase the density of the discretization points and, in this way, to increase the precision. Solution $g(\gamma,z)$ in full interval $-1\leq z\leq 1$ is trivially  obtained from $g_{sym}(\gamma,z)$ (or from $g_{asym}(\gamma,z)$)
using the symmetry (antisymmetry) of  this solution.

Solving Eq. (\ref{eq6}) for $\mu=0.15$ and $\mu=0.5$ numerically, by spline techniques, with number of intervals $N_{\gamma}=N_z=24$,  we reproduced, within three digits, the results shown in the Table I in Ref. \cite{FrePRD14}. Increase of  number of intervals improves the comparison.

\section{Cross-ladder kernel}\label{clk}
As mentioned, the above calculation of the kernel $N$ in the equation (\ref{L6}) was based on the inversion of the kernel $L$ contained in the l.h.-side of the equation (\ref{bsnew}) written symbolically in the form $Lg=Vg$. The recipe of this inversion is universal for any kernel $V$ contained in the r.h.-side of Eq. (\ref{bsnew}).  The result of application of the inverse l.h.-side kernel $L$ to $V$: $N=L^{-1}V$ is given by Eq. (\ref{L7}). 
Therefore, similar calculation of the kernel $N$ can be carried out for any kernel $K$ in the BS equation (\ref{bse}), given by an irreducible Feynman graph
(and, correspondingly, for any kernel $V$ in r.h.-side of Eq. (\ref{bsnew})).
Though, of course, for more complicated $K$ the calculations become more cumbersome.
We illustrate it (more schematically) for the cross-ladder kernel.

\begin{figure}[hbtp]
\centering 
 \includegraphics[width=8cm]{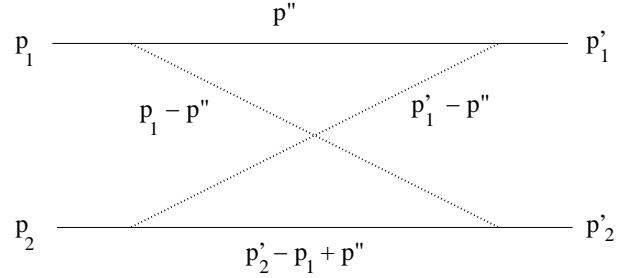}
\caption{Cross-ladder kernel}\label{fig1}
\end{figure}

Feynman graph for the cross-ladder kernel is shown in Fig.~\ref{fig1}.
Corresponding kernel $V^{(CL)}(\gamma,z,\gamma',z')$ in the non-canonical equation (\ref{bsnew}) 
was calculated in \cite{bs2}. We rewrite $V^{(CL)}$ as
\begin{widetext}
\begin{equation}\label{Kcl_1}
V^{(CL)}(\gamma,z,\gamma',z')=-\frac{1}{\pi^2}\alpha^2m^4(1-z^2)^3\int_0^1 y_4(1-y_4)^2dy_4
\int_0^1dy_3\int_0^{1-y_3}dy_2\int_0^{1-y_2-y_3} G \,\eta \,dy_1,
\end{equation}
\end{widetext}
where $\eta=1-y_4[1-(1-y_1-y_3)(y_1+y_3)]$, 
\begin{equation}\label{G}
G=\frac{1}{[\gamma+m^2z^2+\kappa^2(1-z^2)]D^3},
\end{equation}
and $D$ reads:
\begin{eqnarray}\label{Den}
D&=&c_{\gamma}\gamma+c_{\gamma'}\gamma'
+c_{\kappa}\kappa^2+c_{m}m^2+c_{\mu}\mu^2
\nonumber\\
&-&y_4|\tilde{c}|\Bigl[\gamma+m^2z^2+\kappa^2(1-z^2)\Bigr].
\end{eqnarray}
The coefficients $c_{\cdots}$ determining $D$ are given  in  \cite{bs2}.

Making in (\ref{Kcl_1}) the substitution (\ref{L7}), we obtain for  the kernel $N^{CL}(\gamma,z,\gamma',z')$ the expression
\begin{widetext}
\begin{equation}\label{Ncl}
N^{(CL)}(\gamma,z,\gamma',z')=-\frac{1}{\pi^2}\alpha^2m^4(1-z^2)^3\int_0^1 y_4(1-y_4)^2dy_4
\int_0^1dy_3\int_0^{1-y_3}dy_2\int_0^{1-y_2-y_3} n \,\eta \,dy_1,
\end{equation}
\end{widetext}
where
\begin{equation}\label{G1}
n=\frac{1}{2\pi}\int_{-\pi+\epsilon}^{\pi-\epsilon}\frac{d\phi}{(a_1\exp{i\phi}+b_1)^3},
\end{equation}
and
\begin{eqnarray*}
a_1&=&(c_{\gamma}-y_4|\tilde{c}|)\gamma,
\\
 b_1&=&c_{\gamma'} \gamma' -c_{\gamma} (\kappa^2(1-z^2)+m^2z^2)
+c_{\kappa}\kappa^2
\\
&+&c_m m^2+c_{\mu}\mu^2.
\end{eqnarray*}

Like for the ladder kernel, Eq. (\ref{N1}), the result is again given by sum of two integrals (now - 4D integrals over $y_1,y_2,y_3,y_4$) from regular and singular part contributions:
$n=n_{reg}+n_{sing}$.
They are calculated similarly to the ladder case. However, in the ladder case, the values $a$ and $b$ are positive. The condition that the pole $y=-\frac{b(v)}{a(v)}$ is outside the unit cycle means $-\frac{b(v)}{a(v)}<-1$, that is  $b(v)-a(v)>0$. In the cross-ladder case, $a_1$, depending on the values of variable, can be positive or negative, 
that now means: $\left|\frac{b_1}{a_1}\right|>1$, that is  $|b_1|-|a_1|>0$. Hence:
\begin{eqnarray}\label{Jp}
n_{reg}&=&\frac{1}{2\pi i}\int_{\mathcal C}dy\frac{1}{y(a_1y+b_1)^3}=\displaystyle{\frac{1}{b_1^3}}\theta(|b_1|-|a_1|),
\nonumber\\
n_{sing}&=&-\frac{1}{b_1^2}\delta(b_1-a_1).
\end{eqnarray}
The kernel is obtained by substituting in Eqs. (\ref{Jp}) the values $a_1,b_1$  and integrating according to Eq. (\ref{Ncl}).

\section{Conclusion}\label{concl}
We have found the new form of the kernel $N(\gamma,z,\gamma',z')$ of the canonical equation (\ref{L6}):  $g=Ng$,  for the Nakanishi function $g$ for the ladder kernel. This form differs from two forms found previously in Refs. \cite{nak2,FrePRD14}, though all three kernels are equivalent to each other. 
The kernel in the form found in the present paper is expressed via the real functions, does not contain any ambiguities and the method of its calculation can be applied to any kernel given by irreducible Feynman graph. We  outlined this generalization for the cross-ladder BS kernel. For the ladder, the integration is fulfilled analytically and the result for $N$ is given by Eq. (\ref{Knew}) in terms of $\tilde{N}$ determined in Sec. \ref{summary}.
The solutions $g(\gamma,z)$ of Eq. (\ref{L6}), defined in the interval $-1\leq z\leq 1$, are either symmetric relative to $z\to -z$, or antisymmetric.
For both solutions the equation (\ref{L6}) is reduced to Eq. (\ref{eq6}), defined on  the half-interval $0 \leq z\leq 1$. For symmetric solution,  the kernel is defined in Eq. (\ref{eq7}). For antisymmetric one, the kernel is constructed as explained in Sec. \ref{half}.
For the cross-ladder, the kernel  is given by Eq. (\ref{Ncl}). 

The canonical equation (\ref{L6}), in contrast to Eq. (\ref{bsnew}), does not contain the integral in l.h.-side and therefore it is solved easier, its solution is stable relative to the numerical procedure.  The equation (\ref{L6}) is an equation in two variables, with smooth easy calculated kernel, and therefore finding its solution is not more difficult than for the Euclidean BS equation. However, it provides, via Nakanishi function, the BS amplitude in Minkowski space.  Therefore, applications of  these methods to solving the BS equation can give considerable advantages  \cite{ckm_ejpa,ck_trento_09}.
\bigskip

\section*{Acknowledgement}
I am sincerely grateful to Jaume Carbonell kindly provided me his fortran code which I used to solve the equation (\ref{eq6}).


\end{document}